\documentclass[aps,pre,preprint,superscriptaddress]{revtex4-1} 

\usepackage{amsfonts}
\usepackage{amsmath}
\usepackage{amssymb}
\usepackage{graphicx}
\usepackage{mathrsfs} 
\usepackage{color} 

\newcommand{\pd}[2]{\frac{\partial #1}{\partial #2}}

\begin{document}

\title{Spoof surface plasmons guided by narrow grooves}

\author{Ory Schnitzer}
\affiliation{Department of Mathematics, Imperial College London, 180 Queen's Gate, SW7 2AZ, London, UK}

\begin{abstract}
An approximate description of surface waves propagating along periodically grooved surfaces is intuitively developed in the limit where the grooves are narrow relative to the period. Considering acoustic and electromagnetic waves guided by rigid and perfectly conducting gratings, respectively, the wave field is obtained by interrelating elementary approximations obtained in three overlapping spatial domains. Specifically, above the grating and on the scale of the period the grooves are effectively reduced to point resonators characterised by their dimensions as well as the geometry of their apertures. Along with this descriptive physical picture emerges an analytical dispersion relation, which agrees remarkably well with exact calculations and improves on preceding  approximations. Scalings and explicit formulae are obtained by simplifying the theory in three distinguished propagation regimes, namely where the Bloch wavenumber is respectively smaller than, close to, or larger than that corresponding to a groove resonance. Of particular interest is the latter regime where the field within the grooves is resonantly enhanced and the field above the grating is maximally localised, attenuating on a length scale comparable with the period. 
\end{abstract}
\maketitle

\section{Introduction}\label{sec:int}
There are several scenarios where flat interfaces support localised surface waves, e.g. Rayleigh waves guided along an interface of an elastic solid, water waves at fluidic interfaces, and surface-plasmon polaritons at metal--dielectric interfaces (within a frequency interval lying below the surface plasma frequency of the metal). In contrast, a flat boundary of a rigid solid does not normally support surface acoustic waves, nor does a flat boundary of a perfectly conducting metal support surface electromagnetic waves. In such cases, however, surface modes may nevertheless exist if  the interface features transverse stratification or a periodic microstructure. In particular, localised modes propagating along a periodic microstructure, traditionally termed Rayleigh--Bloch waves, have been studied extensively \cite{Brillouin:48,Collin:60} with applications in electromagnetism \cite{Ulrich:73,Sievenpiper:99}, acoustics \cite{Kelders:98} and hydrodynamics \cite{Evans:95}. Ever since a highly influential paper by Pendry \textit{et al.} \cite{Pendry:04}, there has been a remarkable resurgence of interest in Rayleigh--Bloch waves \cite{Garcia:05,Hibbins:05,Abajo:05,Hendry:08,Collin:09,He:11,Hooper:14}. Their paper drew attention to the similarity between electromagnetic surface waves propagating along a periodically microstuctured boundary of a perfect conductor and surface plasmon polaritons. This linkage has enhanced the impact of ideas originally thought pertinent solely to plasmonic materials. Rayleigh--Bloch waves guided by a periodic microstructure are nowadays often referred to (even in acoustics!) as spoof, or designer, surface plasmons. 
 
It is useful to distinguish between two types of spoof plasmons \cite{Christensen:08}. The first, described by Pendry \textit{et al.} \cite{Pendry:04} and later more generally \cite{Garcia:05,Abajo:05,Hendry:08,Collin:09}, is an electromagnetic surface wave guided by a perfectly conducting surface with holes in it. Typical dispersion surfaces  bifurcate with increasing Bloch wavenumber from the light cone to lower frequencies, particularly below the cut-off frequency of the hole waveguides. Since the latter frequency scales inversely with the linear dimensions of the holes' cross section, which are bounded by the periodicity, the wavelength cannot be large compared to the periodicity unless the holes are filled with a high-index material. The present paper is concerned with a second class of spoof plasmons, where the microstructure consists of holes or grooves functioning as waveguide-like elements that do not have a cut-off frequency. This is generally the case for acoustic waves in rigid waveguides \cite{Kelders:98,Christensen:08}, electromagnetic waves penetrating into perfectly conducting grooves (magnetic field parallel to the grooves) \cite{Collin:60,Garcia:05}, as well as for some thin grating structures \cite{Huidobro:14,Shen:14}. In these cases the wave field is free to propagate up and down the holes or grooves; the resonance frequencies of these waveguides scale inversely with their length and constitute upper bounds on spoof-plasmon frequencies of respective order. Such spoof plasmons can accordingly be tuned to low frequencies by simply lengthening the holes/grooves (spiral geometries allow a compact design \cite{Huidobro:14,Hooper:14}). 

In the absence of a cut-off frequency, the holes/grooves can in principle be made arbitrarily narrow compared to the periodicity, promoting the excitation of waveguide modes and hence the field enhancement within them. Specifically, as the waveguide width decreases the $n$th spoof-plasmon eigenfrequency approaches (for fixed Bloch wavenumber) the lower between the light-line frequency and the respective waveguide-resonance frequency; a naive leading-order approximation of the dispersion curves in this limit is simply the piecewise-linear functions thus formed. It is clear, however, that the limit of narrow waveguides is singular. Namely, these piecewise-linear curves can never be realised; their linear segments respectively correspond to the modes of the waveguides and bulk, rather than true surface modes. Accordingly, in order to characterise the surface modes in the limit of narrow holes/grooves it is necessary to determine the leading deviations of the dispersion curves from the singular ones.

In this paper we address this challenge by developing a consistent and physically intuitive theory of Rayleigh--Bloch waves guided by a periodically grooved surface, in the limit where the grooves are narrow compared to the period. We formulate the eigenvalue problem in section \ref{sec:form}, in the case where the waves are acoustic or electromagnetic and the grating is respectively rigid or perfectly conducting. Rather than attempting to reduce an exact formulation, the theory is developed in section \ref{sec:narrow} directly from the governing equations, which are separately simplified in three overlapping spatial domains. In each of these domains we obtain an elementary approximation, closed by requiring consistency where any two of these regions overlap. Of course, this is found to be possible only for certain combinations of frequency and Bloch wavenumber, thus giving rise to a dispersion relation which we show to be in excellent agreement with exact numerical solutions. In section \ref{sec:study} we further simplify the theory in three distinguished propagation regimes corresponding to overlapping asymptotic intervals of the Bloch wavenumber. We thereby derive scalings and explicit asymptotic formulae for eigenfrequencies, localisation length scales and enhancement factors. We conclude in section \ref{sec:conc} by discussing the relation between our work and existing approximate theories. 

\section{Formulation}\label{sec:form}
\begin{figure}[t]
\begin{center}
\includegraphics[scale=0.65]{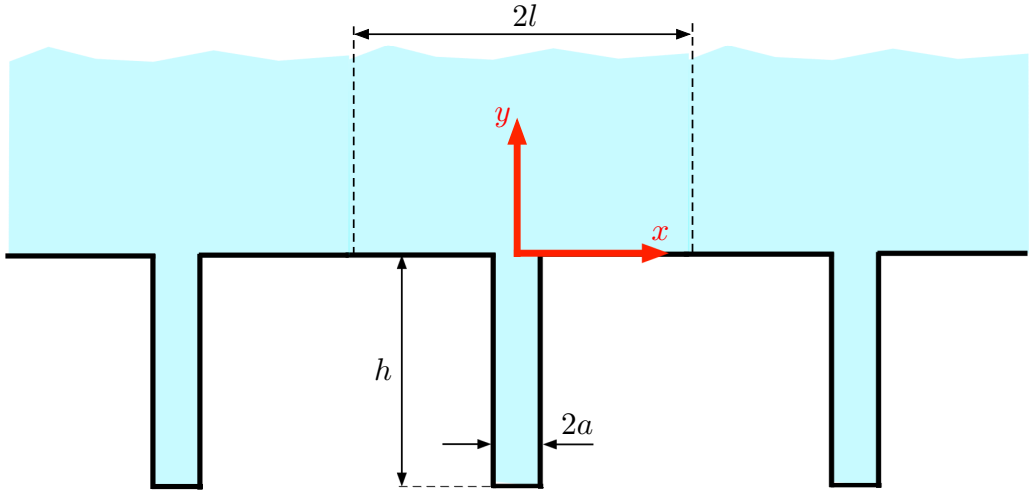}
\caption{Schematic of the two-dimensional rectangular grating. }
\label{fig:schematic}
\end{center}
\end{figure}

Consider a surface patterned with a $2l$-periodic array of rectangular grooves of width $2a$ and height $h$, as shown in figure \ref{fig:schematic}. Let $\varphi(x,y)\exp(-i\omega t)$ be the velocity potential in the acoustic case and the out-of-plane magnetic field in the electromagnetic case, where $\omega$ denotes angular frequency, $t$ is time, and $(x,y)$ are the in-plane Cartesian coordinates $(x,y)$ defined in figure \ref{fig:schematic}. Outside the surface the wave field $\varphi$ obeys the reduced wave equation 
\begin{equation}\label{master}
\nabla^2\varphi+k^2\varphi=0,
\end{equation}
where $k=\omega/c$ is the wavenumber, $c$ being the speed of sound or light in the acoustic and electromagnetic cases, respectively. At the surface, which is assumed to be acoustically rigid in the acoustic case and perfectly conducting in the electromagnetic case, $\varphi$ satisfies a Neumann boundary condition,
\begin{equation}\label{neumann}
\pd{\varphi}{n}=0,
\end{equation}
where $\partial{}/\partial{n}$ denotes differentiation normal to the surface. 
To constitute a surface mode, the field $\varphi$, defined up to a multiplicative constant, must attenuate at large distances from the surface,
\begin{equation}\label{decay}
\varphi\to0 \quad \text{as} \quad y\to\infty,
\end{equation}
and be Bloch periodic, i.e.,
\begin{equation}\label{bloch}
e^{-i\beta x} \varphi(x,y)  \text{ is } 2l \text{ periodic in } x,
\end{equation}
where the Bloch wavenumber $\beta$ is real valued. Accordingly, attention is restricted to a single unit-cell, say $x\in[-l,l)$, placing the origin $(x,y)=(0,0)$ at the centre of the interface of an arbitrarily chosen groove. 

The above formulation defines an eigenvalue problem for the wavenumbers $k(\beta)$ and corresponding eigenfunctions $\varphi$. Given the periodicity of the grating and time-reversal symmetry it is sufficient to consider Bloch wavenumbers in the reduced Brillouin zone $0<\beta<\pi/(2l)$. With $\beta$ in that range, a Fourier-series representation of $\varphi$ suggests the asymptotic behaviour
\begin{equation}\label{asymptotic}
\varphi \sim \text{const} \times e^{-y\sqrt{\beta^2-k^2}} \quad \text{as} \quad y\to\infty,
\end{equation}
which is consistent with \eqref{decay} if $|\beta|>k$. It is therefore sufficient to consider Bloch wavenumbers in the range $k<\beta<\pi/(2l)$. 

\section{Narrow grooves}\label{sec:narrow}
In what follows we assume that $ka\ll kl = O(1)$, namely, that the grooves are narrow compared to the period, the latter being comparable to the wavelength. More specifically, the range of frequencies in which we expect to find surface waves, $kh = O(1)$, is determined by the height of the grooves. As noted above, in general we assume that $l$ and $h$ are comparable, though as we go along we shall note where and how our theory simplifies for $h\gg l $, in which case the period is subwavelength, $kl\ll1$.
\subsection{Within the groove}
\begin{figure}[b]
\begin{center}
\includegraphics[scale=0.6]{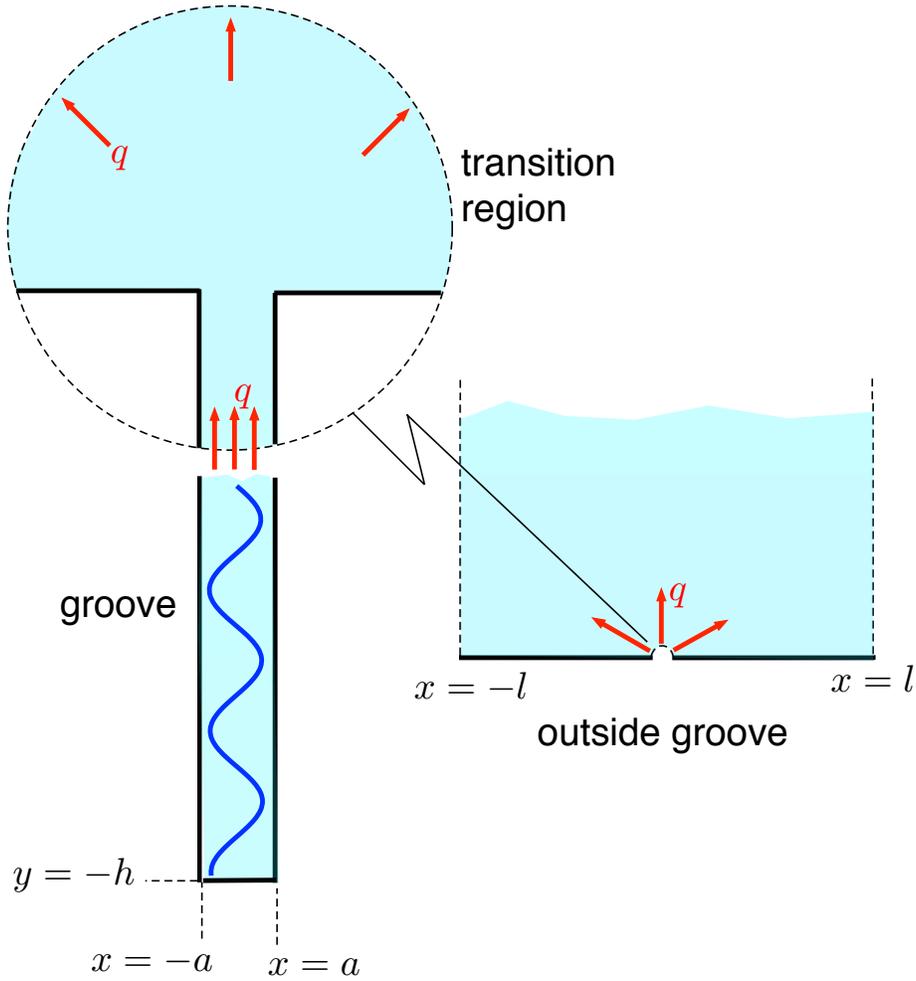}
\caption{An approximate theory in the limit $a/l\to0$ is sought by conceptually decomposing a representative unit cell into three regions.}
\label{fig:regions}
\end{center}
\end{figure}

We start by constructing an approximation of the wave field $\varphi$ that is valid within the groove, at distances $\gg a$ from its aperture. Owing to the narrowness of the groove, $a/h\ll 1$, along with the Neumann conditions \eqref{neumann} on its boundaries $x=\pm a$, in this region $\varphi$ is approximately a function of $y$ alone, i.e.~$\varphi\approx u(y)$. The reduced wave equation \eqref{master}  becomes $u''+k^2u=0$, while condition \eqref{neumann}, applied to the bottom end of the groove, implies $u'(-h)=0$. Hence
\begin{equation}\label{groove approx}
\varphi \approx C\left[\cos(ky)-\tan(kh)\sin(ky)\right], \quad a\ll-y< h, 
\end{equation}
where $C$ is a constant. Since $\varphi$ is unique up to a multiplicative constant, $C$ can be chosen arbitrarily. We shall require \eqref{groove approx} to be consistent with complementary approximations of $\varphi$ outside the groove and in the vicinity of the aperture (see figure \ref{fig:regions}). This procedure will give rise to an additional constraint on \eqref{groove approx}, thus yielding a dispersion relation for $k(\beta)$. With  anticipation, we note that close to the groove opening, \eqref{groove approx} becomes
\begin{equation}\label{groove ent overlap}
\varphi \approx C + \frac{q}{2a}y, \quad a\ll  -y \ll h,
\end{equation}
where 
\begin{equation}\label{flux}
q = -2aCk\tan(kh)
\end{equation}
is, in the acoustics case, the net flux through the groove. In the electromagnetic case $q$ is proportional to the voltage across the aperture. 

\subsection{Outside the groove} 
Consider next the region of the unit cell outside the groove. Since $a/l\ll1$, on the scale of the period the  aperture shrinks to the origin $(x,y)=0$. Nevertheless, we expect the finite flux $q$ to emanate from this point into the external domain (the flux is conserved while passing through the subwavelength aperture of the groove). 
Thus, when resolved on the scale of the period (or any scale $\gg a$), $\varphi$ appears to possess the singularity
\begin{equation}\label{log sing}
\varphi \sim \frac{q}{\pi}\ln \frac{\sqrt{x^2+y^2}}{l} \quad \text{as} \quad x^2+y^2\to0, \quad y\ge 0.
\end{equation}
An approximation for the external potential can then be obtained by solving \eqref{master} in conjunction with the Bloch condition \eqref{bloch}, the Neumann condition \eqref{neumann} applied at $y=0$ ($-l<x<l$), the attenuation condition \eqref{decay}, and the singular asymptotics \eqref{log sing}. Fourier methods readily yield a formal solution
\begin{equation}\label{fourier sol}
\varphi \approx -\frac{q}{2l}e^{i\beta x}\sum_{m=-\infty}^{\infty}\frac{e^{-y\sqrt{\left(\beta+m\pi/l\right)^2-k^2}+im\pi x/l}}{\sqrt{\left(\beta+m\pi/l\right)^2-k^2}}, \quad \quad \sqrt{x^2+y^2}\gg a , \quad y\ge 0,
\end{equation}
namely a quasi-periodic Green function of the two-dimensional Helmholtz equation, whose magnitude is adjusted to satisfy \eqref{log sing}. Our interest is not in the details of this solution. Rather, we require its behaviour near the origin, so that it could be related to that within the groove. Setting $y=0$ in \eqref{fourier sol}, then taking the limit $x\to 0$, one finds \cite{Linton:10}
\begin{equation}\label{out ent overlap}
\varphi \approx \frac{q}{\pi}\ln \frac{\sqrt{x^2+y^2}}{l} - \frac{q}{2}\left[\frac{1}{\sqrt{\beta^2l^2-k^2l^2}}\right. \left. -\frac{2}{\pi}\ln \pi +S(\beta l, k l)\right], \quad a\ll \sqrt{x^2+y^2}\ll l,
\end{equation}
where the sum
\begin{equation}\label{sum}
S(\beta l, k l) = {\sum_{\substack{m=-\infty \\ m\neq 0}}^{\infty}}\left(\frac{1}{\sqrt{(\beta l +m\pi)^2-k^2l^2}}-\frac{1}{|m|\pi}\right)
\end{equation}
is easy to compute. In fact, in the subwavelength regime, where $kl\ll1$, this sum can be evaluated in closed form as
\begin{equation}\label{sum sw}
S(\beta l, k l) \approx S(\beta l, 0) = -\frac{2\gamma}{\pi}-\frac{1}{\pi}\left[\psi\left(1-\frac{\beta l}{\pi}\right)+\psi\left(1+\frac{\beta l }{\pi}\right)\right],
\end{equation}
where $\psi$ is the digamma function and $\gamma \approx 0.5772$ is the Euler-Mascheroni constant.

\subsection{Transition region} 
Approximations \eqref{groove ent overlap} and \eqref{out ent overlap} hold, respectively within and outside the groove, at distances from the origin that are much larger than $a$ and at the same time much smaller than $l$ (and $h$). As these regions do not overlap, it is unsurprising that in general it is impossible to make \eqref{groove ent overlap} and \eqref{out ent overlap} coincide. As schematically shown in figure \ref{fig:regions}, to smoothly join these approximations it is necessary to consider a transition region in the vicinity of the aperture. On this subwavelength scale \eqref{master} reduces to Laplace's equation,
\begin{equation}\label{transition}
\nabla^2 \varphi \approx 0, \quad \sqrt{x^2+y^2}\ll l,
\end{equation}
and the boundary, where Neumann condition \eqref{neumann} applies, is effectively that of an infinite groove interfacing an unbounded half-space (see figure \ref{fig:regions}). 
We seek a solution of \eqref{transition} in the latter geometry which for large negative $y$ coincides with \eqref{groove ent overlap} and for large $x^2+y^2$ ($y>0$) coincides with \eqref{out ent overlap}. 
\begin{figure}[b]
\begin{center}
\includegraphics[scale=0.8]{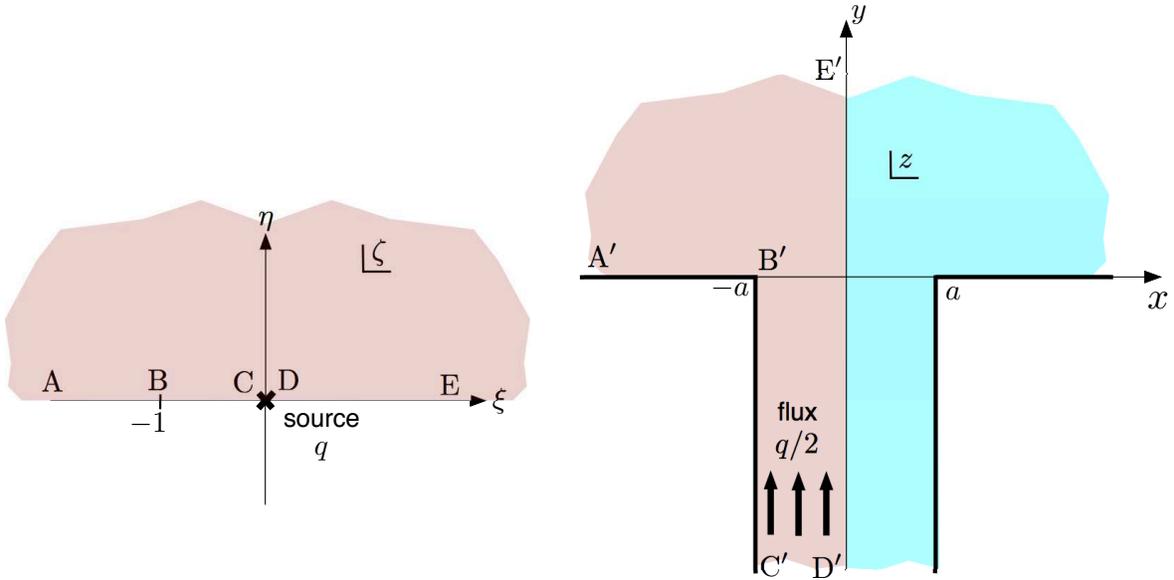}
\caption{The left half $(x<0)$ of the transition region (see figure \ref{fig:regions}) is conformally mapped to the upper half plane of an auxiliary $\zeta$ plane.}
\label{fig:mapping}
\end{center}
\end{figure}

Up to an additive constant, the requisite potential is determined by the net flux $q$, which corresponds to the non-constant terms in \eqref{groove ent overlap} and \eqref{out ent overlap}.  Thus, obtaining this solution would in turn determine the difference between the constant terms in \eqref{groove ent overlap} and \eqref{out ent overlap} (when normalised by $q$ this is a purely geometric parameter \cite{Taylor:73}). To this end, let $z = x+iy$ and consider the conformal mapping \cite{Brown:book}
\begin{equation}\label{mapping}
z/a = \frac{2i}{\pi}\left(1+\zeta\right)^{1/2}+\frac{i}{\pi}\log\frac{(1+\zeta)^{1/2}-1}{(1+\zeta)^{1/2}+1}
\end{equation}
from the upper half-plane of an auxiliary complex plane $\zeta=\xi+i\eta$ to the left half $(x<0)$ of the solution domain, in the manner shown in figure \ref{fig:mapping}. (The logarithm is the continuation of the natural logarithm with a branch cut along the negative imaginary axis.) In terms of the auxiliary variable $\zeta$, the potential in the transition region is determined as 
\begin{equation}\label{zeta sol}
\varphi/q \approx \frac{1}{2\pi}\ln|\zeta|+\alpha,
\end{equation}
up to the multiplicative constant $q$ (which follows from \eqref{flux} and the arbitrary choice of the constant $C$), and the real constant $\alpha$, which we shall determine by relating \eqref{zeta sol} to the solutions inside and outside the groove. 
To this end, we note the following asymptotic limits of the mapping \eqref{mapping}, 
\begin{gather}\label{mapping asym}
z/a\sim \frac{i}{\pi}\log\zeta+\frac{2i}{\pi}(1-\ln 2) + o(1) \quad \text{as} \quad \zeta\to0,\\
z/a\sim \frac{2i}{\pi}\zeta^{1/2} \quad \text{as} \quad \zeta\to\infty.
\end{gather}
Then, from \eqref{zeta sol}, we find the ``overlap'' approximations
\begin{equation}\label{ent overlap groove}
\varphi/q \approx \frac{y}{2a}-\frac{1}{\pi}(1-\ln 2)+\alpha, \quad a\ll  -y \ll h
\end{equation}
and
\begin{equation}\label{ent overlap out}
\varphi/q \approx \frac{1}{\pi}\ln\frac{\sqrt{x^2+y^2}}{l}+\frac{1}{\pi}\ln\frac{\pi l}{2a}+\alpha, \quad a \ll  \sqrt{x^2+y^2} \ll l.
\end{equation}

\subsection{Dispersion relation} 
Comparing \eqref{ent overlap groove} with \eqref{groove ent overlap}, and \eqref{ent overlap out} with \eqref{out ent overlap}, and using \eqref{flux}, we find 
\begin{equation}\label{dispersion}
\frac{1}{ka\tan(kh)}\approx \frac{1}{\sqrt{(\beta l)^2-(k l )^2}}+\frac{2}{\pi}\left(1+\ln\frac{l}{4a}\right) + S(\beta l , k l)
\end{equation}
and $\alpha = (1-\ln 2)/\pi-1/[2k a\, \tan(k h)]$. The dispersion relation \eqref{dispersion} constitutes a key result of this paper, as do the corresponding approximations found for the eigenmode $\varphi$ in the different regions. In the next section, we shall show how \eqref{dispersion} and the latter approximations of $\varphi$ simplify in different asymptotic intervals of the Bloch wavenumber $\beta$.

\begin{figure}[h]
\begin{center}
\includegraphics[scale=0.5]{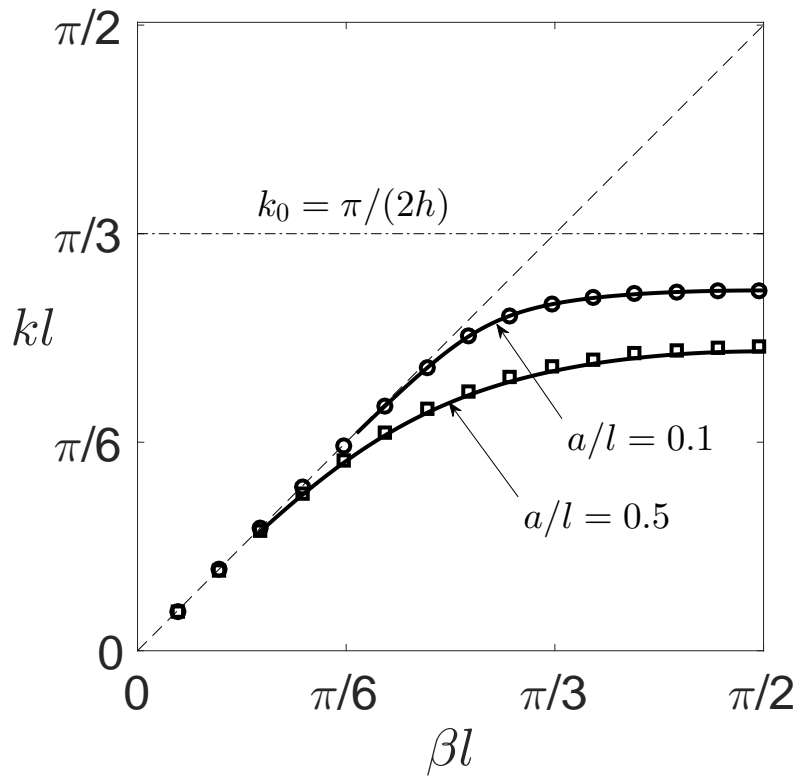}
\caption{Dispersion curves for $h/l=1.5$, for the two indicated values of $a/l$. Solid lines: solutions of the dispersion relation \eqref{dispersion}; symbols: `exact' numerical solutions. The diagonal dashed line and horizontal dash-dotted line respectively indicate the light line and the fundamental groove resonance.}
\label{fig:com1}
\end{center}
\end{figure}

\begin{figure}[h]
\begin{center}
\includegraphics[scale=0.6]{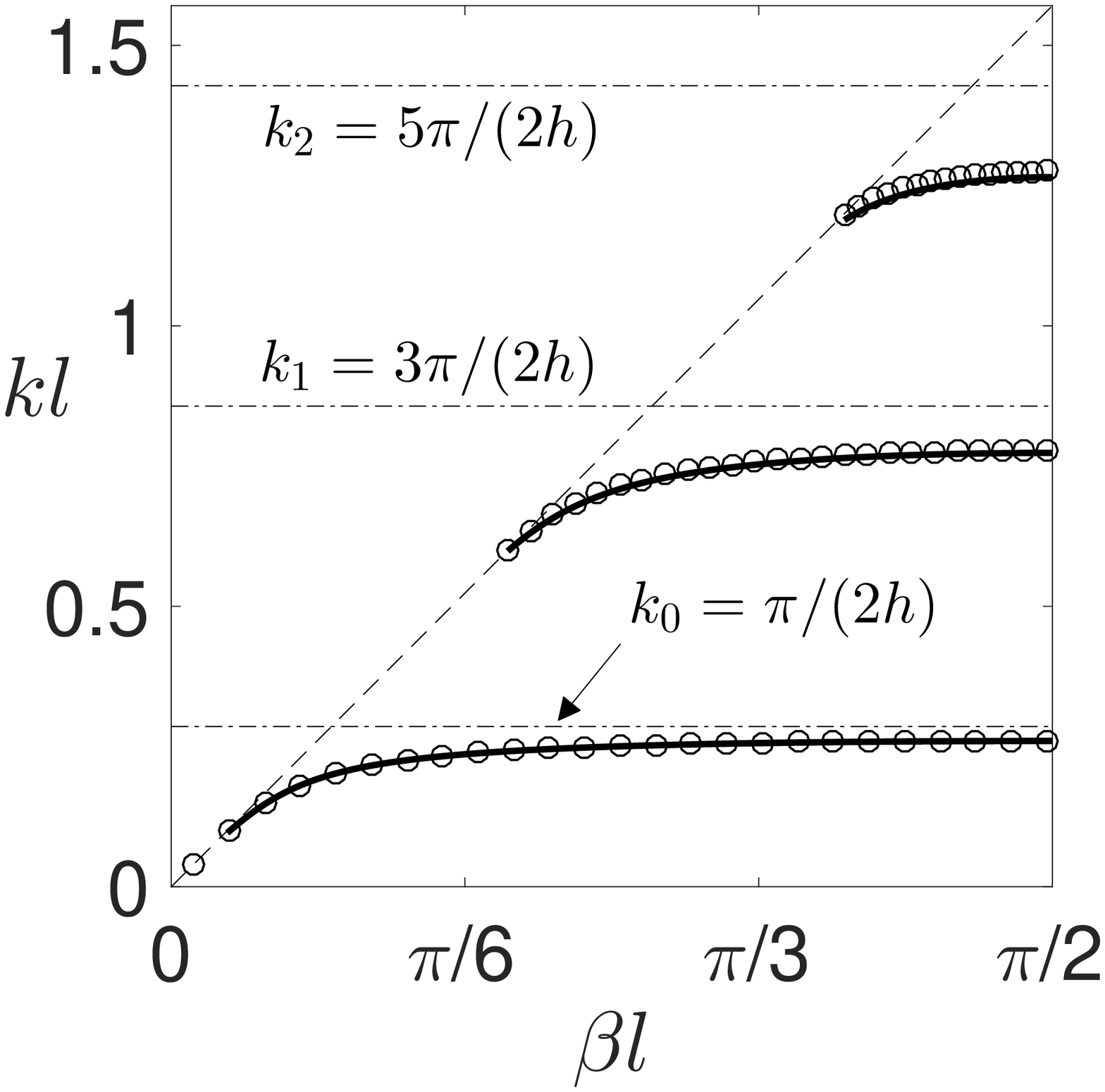}
\caption{Dispersion curves for $h/l=5.5$ and $a/l=0.5$. Solid lines: solutions of the dispersion relation \eqref{dispersion}; symbols: `exact' numerical solutions. The diagonal dashed line and horizontal dash-dotted lines respectively indicate the light line and the first three groove resonances.}
\label{fig:com2}
\end{center}
\end{figure}

In figures \ref{fig:com1} and \ref{fig:com2}, dispersion curves $k(\beta)$ found by solving \eqref{dispersion} are depicted by the black solid lines, for various values of the geometric parameters $h/l$ and $a/l$. The diagonal dashed ``light'' line depicts $k=\beta$, and the horizontal dash-dotted lines mark the groove resonances, 
\begin{equation}\label{kn}
k_n =\left(\frac{1}{2}+n\right)\frac{\pi}{h}, \quad n=0,1,2,\ldots,
\end{equation}
which are discussed in the next section. The fundamental resonance $k_0$ corresponds to a wavelength four times the groove height $h$. In figure \ref{fig:com1}, where $h/l=1.5$,  only the fundamental $k_0$ lies below $k = \pi/(2l)$, at which frequency the light line intersects the zone boundary. There is just one surface mode, with $k$ bounded from above by the minimum of $\beta$ and $k_0$. With increasing $h$, additional resonant wavenumbers \eqref{kn} drop below $\pi/(2l)$. For each of these ($n=0,1,2,\ldots$) there is one surface mode for which $k$ approaches the smaller of $\beta$ and $k_n$ as $a/l\to0$. This is demonstrated in figure \ref{fig:com2}, where $h/l=5.5$ and there are three branches. (There may also be ``weakly guided'' modes which as $a/l\to0$ approach the light line for all $\beta$. This is certainly the case when $h/l<1$ in which case there are zero groove resonances below $k=\pi/(2l)$.) 

In order to assess the approximate dispersion relation \eqref{dispersion}, we have solved the problem formulated in section \ref{sec:form} exactly using a semi-numerical mode-matching method. Since the scheme we use is standard and similar to those discussed elsewhere \cite{Mitra:71,Kelders:98,Abajo:05,Hendry:08}, we omit details here. In figures \ref{fig:com1} and \ref{fig:com2}, the numerical predictions are marked by the symbols and remarkable agreement with the approximate theory is observed; surprisingly, the agreement is quite good even  for moderate values of $a/l$. In section \ref{sec:conc} we shall revisit the dependence of the error upon $a/l$ in relation to existing approximations in the literature. 

\section{Three propagation regimes} 
\label{sec:study}
The approximate theory developed in the preceding section holds for arbitrary $\beta$. To the same order of approximation in the limit $ka\to0$, however, further simplifications are permissible in specified intervals. This is hinted by the survival of the small parameter $ka$  in the approximate dispersion relation \eqref{dispersion}. In what follows we discuss and derive scalings and explicit formula pertaining to three distinguished limits of \eqref{dispersion}, which respectively correspond to three distinct regimes of surface-wave propagation. In particular, these regimes differ in terms of localisation of the wavefield to the grating, the excitation of groove resonances and the concomitant field enhancement within the grooves. As a preliminary step, we note the following general estimates, which we shall subsequently sharpen in each of the propagation regimes. First, it readily follows from \eqref{asymptotic} that the length scale on which the wavefield attenuates away from the grating is 
\begin{equation}\label{length}
L = O\left(\frac{1}{\sqrt{\beta^2-k^2}}\right). 
\end{equation}
Since $k<\beta<\pi/(2l)$, $L$ is at least comparable to the period, the localisation becoming weaker as the wavenumber $k$ approaches the light line. Next, we obtain an estimate for the enhancement of the field in the groove relative to the field in the outer region. From \eqref{groove approx}, the field within the groove is $O(C \tan(kh))$, where the tangent is retained since it becomes large close to a groove resonance \eqref{kn}. From \eqref{fourier sol},  we estimate the field in the outer region as $O(q/\sqrt{\beta^2l^2-k^2l^2})$. Using \eqref{flux}, we obtain a preliminary estimate of the field enhancement,
\begin{equation}\label{enhancement}
A =O\left(\frac{l/{L}}{ka}\right),
\end{equation}
revealing a linkage between localisation and field enhancement.

\subsection{Strongly guided waves}
By strongly guided surface waves we mean $L=O(l)$, namely that $k$ is not close to $\beta$. In that case, the right hand side of \eqref{dispersion} is of order unity, implying that $\tan(kh)$ is $O(1/(ka))$ large. 
Noting that [cf.~\eqref{kn}] $\tan(kh) \sim h^{-1}(k_n-k)^{-1}$ as $k\to k_n$, such a balance is possible only for $(k_n-k)h=O(ka)$. An explicit formula is readily extracted from \eqref{dispersion}: 
\begin{equation}\label{strong}
\frac{k}{k_n}\approx 1 -\frac{a}{h}\left[\frac{1}{\sqrt{(\beta l )^2-(k_nl)^2}}+\frac{2}{\pi}\left(1+\ln\frac{l}{4a}\right)+ S(\beta l , k_n l)\right].
\end{equation}
If $n$ is sufficiently small and $h$ sufficiently large such that $k_nl\ll1$, then $kl\ll1$, i.e., the period is subwavelength. Approximation \eqref{strong} can then be further simplified:
\begin{equation}\label{strong sw}
\frac{k}{k_n}\approx 1-\frac{a}{h}\left[\frac{1}{\beta l }+\frac{2}{\pi}\left(1+\ln\frac{l}{4a}\right)+ S(\beta l , 0)\right].
\end{equation}
In particular, at the band edge \eqref{strong sw} becomes 
\begin{equation}\label{strong sw edge}
\frac{k}{k_n}\approx 1-\frac{2a}{\pi h}\left(\ln \frac{l}{a}+1 \right), \quad \beta l = \pi/2, \quad kl\ll1,
\end{equation}
where we used \eqref{sum sw} to show that $S(\pi/2,0)=(2/\pi)(\ln 4- 1)$. 

The regime of strongly guided waves is associated with the excitation of a groove resonance. Indeed, we see from \eqref{groove approx} that for $k\approx k_n$ the field within the groove is dominated by a resonant mode proportional to $\sin(k_ny)$, which vanishes in the limit $y\to0$ and satisfies the Neumann condition at $y=-h$. 
Using \eqref{enhancement} we find $A=O(k^{-1}a^{-1})$, namely that the field in the groove is strongly enhanced. 

\subsection{Weakly guided waves}
On the other extreme, the interval of Bloch wavenumbers $\beta$ where the dispersion curve begins to bifurcate from the light corresponds to a regime of weakly guided surface waves. Thus, for $k$ close to $\beta$ but not to $k_n$, we extract from \eqref{dispersion} an explicit approximation,
\begin{equation}\label{weak}
k/\beta   \approx 1-  \frac{1}{2}   \frac{a^2}{l^2}\tan^2(\beta h) .
\end{equation}
In particular we see that $\sqrt{\beta^2-k^2} = O(\beta a /l)$, and using $\beta=O(1/l)$, we find a large attenuation length $L/l=O(l/a)$. The resonant mode in the groove is not excited in this case, and it follows from \eqref{enhancement} that $A=O(k^{-1}l^{-1})$.

\subsection{An intermediate regime}
The transition between the above two regimes takes place over an intermediate regime where both $\beta$ and $k$ are close to $k_n$. In this case, inspecting \eqref{dispersion} suggests the leading-order balance
\begin{equation}\label{int1}
ka\tan(kh)\approx \sqrt{(\beta l)^2-(k l )^2},
\end{equation}
which is the same as Eq.~(14) in \cite{Garcia:05} (see section \ref{sec:conc}). The reduced dispersion relation \eqref{int1} incorporates approximation \eqref{weak} for weakly guided waves. In the intermediate regime, \eqref{int1} further simplifies to 
\begin{equation}\label{int2}
(k_nl-kl)\sqrt{\beta l -k l} \approx \frac{a}{h}\sqrt{\frac{k_n l}{{2}}},
\end{equation}
implying the scaling of the intermediate regime  
\begin{equation}\label{int3}
k_nl-kl,\, \beta l-k_nl = O\left(\frac{a^{2/3}}{l^{2/3}}\right).
\end{equation}
Then, from \eqref{length} and \eqref{enhancement} we find $L/l=O(l^{1/3}a^{-1/3})$ and $A=O(a^{-2/3}l^{-1/3}k^{-1})$; using $k\approx k_n = O(1/h)$ this estimate becomes $A=O(a^{-2/3}h^{2/3})$. Unsurprisingly, the localisation and enhancement are intermediate between the strong and weak guiding regimes. 

The reduced dispersion relation \eqref{int1} (or \eqref{int2}) provides only a gross leading-order deviation of $k$ from $k_n$. Indeed, asymptotic analysis of  \eqref{dispersion} in the intermediate limit \eqref{int3} reveals a substantial $O(a^{1/3}l^{-1/3})$ relative error, which needs to be resolved in order for the approximation in the intermediate regime to smoothly connect with that in the strong guiding regime. Thus, writing
\begin{equation}\label{int beta}
\beta/k_n = 1 +  \left(\frac{a}{l}\right)^{2/3}\delta, \quad \delta=O(1),
\end{equation}
we find from \eqref{dispersion} the improved approximation
\begin{equation}\label{int asym}
k/k_n  \approx 1 +  \left(\frac{a}{l}\right)^{2/3}\nu +  \frac{a}{l}\chi,
\end{equation}
where
\begin{equation}\label{int asym2}
\delta = \nu + \frac{1}{2k_n^2h^2\nu^2}, \quad \chi = -\frac{l}{h\left(1+ k_n^2h^2|\nu|^3\right)}\left[\frac{2}{\pi}\left(1+\ln\frac{l}{4a}\right) + S(k_n l,k_n l)\right]. 
\end{equation}

\begin{figure}[t]
\begin{center}
\includegraphics[scale=0.5]{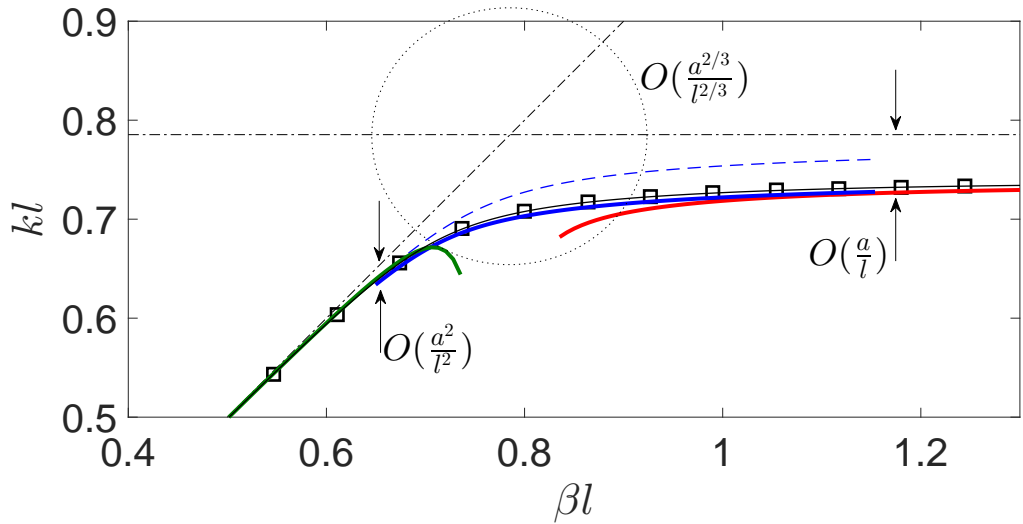}
\caption{Dispersion curve for $h=2l$ and $a/l=0.05$, in the vicinity of $\beta=k_0=\pi/(2h)$. Black solid line: solution of the dispersion relation \eqref{dispersion}; red solid line: approximation \eqref{strong} for strongly guided waves; green solid line: approximation \eqref{weak} for weakly guided waves; blue dashed and solid lines: respectively two and three terms of the intermediate-regime approximation \eqref{int asym}; symbols: `exact' numerical data. The diagonal and horizontal dashed lines respectively indicate the light line and the fundamental groove resonance. The scalings of the three propagation regimes are schematically depicted.}
\label{fig:coal}
\end{center}
\end{figure}

The validity of the approximations found for the different propagation regimes is demonstrated in figure \ref{fig:coal}, showing, for $h=2l$ and $a/l=0.05$, the dispersion curve of the fundamental mode in the vicinity of the intersection of the light line $k=\beta$ and the resonant wavenumber $k_0=\pi/(2h)$. In this figure, the different approximations \eqref{strong}, \eqref{weak}, and \eqref{int asym} are compared with the uniformly valid dispersion relation \eqref{dispersion}, along with exact numerical data. Excellent agreement is found in the respective regimes. 

\section{Concluding remarks}\label{sec:conc}
\begin{figure}[t]
\begin{center}
\includegraphics[scale=0.46]{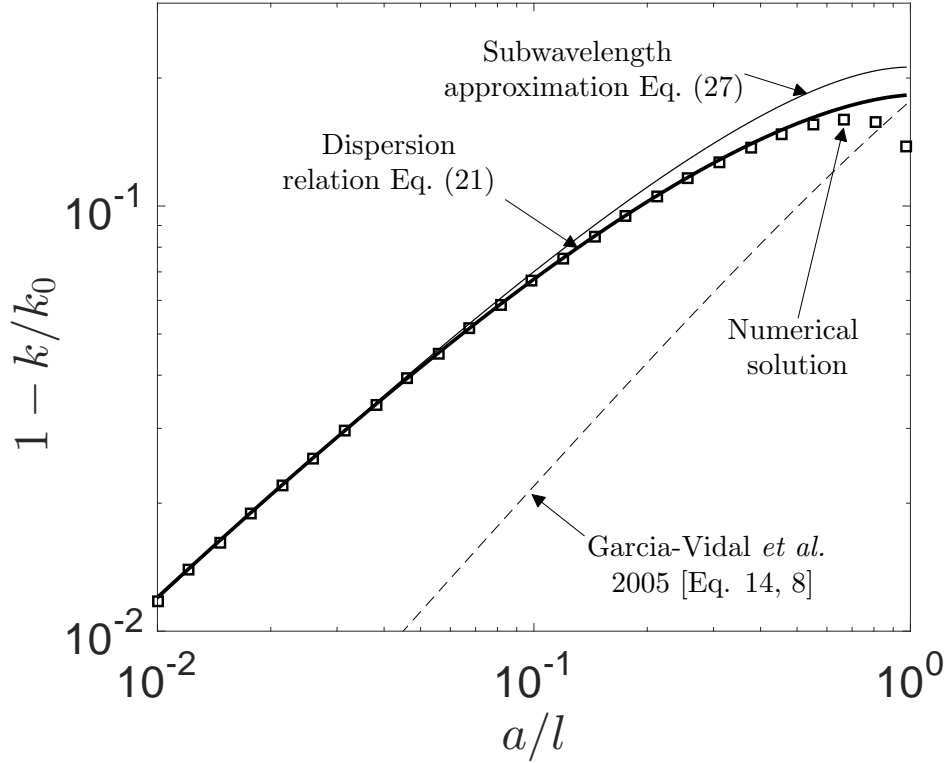}
\caption{Relative deviation of the wavenumber $k$ from the resonant wavenumber $k_0$  as a function of $a/l$ (for $h/l=3$, at the boundary of the Brillouin zone, $\beta l =\pi/2$). Thick solid line: solution of the dispersion relation \eqref{dispersion}; thin solid line: subwavelength approximation \eqref{strong sw edge}; symbols: `exact' numerical solution. The dashed line shows the solution of the reduced dispersion relation \eqref{int1}, which is the narrow-groove approximation given by Garcia-Vidal \textit{et al.} [Eq.~14, 8].}
\label{fig:asym}
\end{center}
\end{figure}
Shortly following the influential paper by Pendry \text{et al.} \cite{Pendry:04}, the same authors (in reversed order) published a more detailed paper \cite{Garcia:05} which includes an analysis of the grooves configuration considered herein.  The analysis in \cite{Garcia:05} assumes, \textit{inter alia}, that the field within the grooves can be represented by a single waveguide mode, all the way up to the aperture. While this assumption is widespread in the literature on spoof plasmons, it contrasts the present study wherein the subwavelength details of the field in the close vicinity of the aperture are found to be important. To be specific we note that Garcia-Vidal \textit{et al.} \cite{Garcia:05} provide their Eq.~(14) --- equivalent to our Eq.~\eqref{int1} --- as a general approximate dispersion relation in the limit of narrow grooves. We have seen in section \ref{sec:study}, however, that while \eqref{int1} correctly describes the weakly guiding regime, it provides only a gross leading-order approximation in the intermediate regime and does not hold (in any asymptotic sense) in the strong guiding regime. In fact, extrapolating \eqref{int1} into the latter regime suggests a deviation from the groove resonance scaling like $k_nl -kl=O(a/l)$, whereas the actual scaling is $(a/l)\log(a/l)$ --- in our theory the leading deviation in the strong guiding regime is given by the explicit formulae \eqref{strong}--\eqref{strong sw edge}. The asymptotic validity of our theory, in contrast to that in \cite{Garcia:05}, is clearly demonstrated in figure \ref{fig:asym}.

Recently, Eremntchouk \textit{et al.} \cite{Erementchouk:16} attempted a more systematic approximation for narrow grooves by intricately reducing an exact mode-matching formulation. These authors identify the strongly guided regime of developed spoof plasmons, and that the standard approximations employed in the literature do not hold in this domain. They focus their efforts on the latter regime and derive an explicit expression for the leading deviation $kl-k_0l$ (they consider only the fundamental surface mode). Their approximation can be shown to agree with our Eq.~\eqref{strong} at $O((a/l)\ln l/a)$ but not $O(a/l)$ --- these successive orders are practically inseparable. In particular, in the limit where the period is subwavelength ($kl\ll1$), their approximation at the edge of the Brillouin zone reduces to 
$
1-k/k_0\approx \frac{2a}{\pi h}\left(\ln\frac{l}{a}+\frac{5}{2}-\ln{2\pi}\right).
$
This differs at $O(a/h)$ from our \eqref{strong sw edge}, which is validated in figure \ref{fig:asym}. Perhaps the discrepancy can be traced to Eq.~(3.11) in \cite{Erementchouk:16},  where a sum similar to the one in our Eq.~\eqref{fourier sol} is approximated.  While the authors claim that ``the main terms are kept'', their approximation is actually only logarithmically accurate. 

To conclude, the method devised in this paper for studying spoof plasmons is simple and physically intuitive, and we anticipate that it can be adopted to study a wide range of propagation and excitation problems in acoustics and photonics involving small holes or grooves \cite{Genet:07}. Our approach, which can be made rigorous using the method of matched asymptotic expansions \cite{Hinch:91,Crighton:12}, follows directly from the governing equations and accordingly is not limited to configurations where an exact formulation is available. Indeed, narrow grooves and holes of arbitrary centreline and (subwavelength) cross section are generally amenable to a quasi-one-dimensional analysis; analysis of the vicinity of an arbitrary aperture can typically be reduced to the calculation, either analytically or numerically, of a small number of purely geometric parameters; and in the external regions outside the microstructured surfaces the holes and grooves can be represented by lumped point-size resonators.  

\bibliography{refs.bib}

\end{document}